\documentclass[journal,10pt]{IEEEtran}
\IEEEoverridecommandlockouts
\usepackage{cite}
\usepackage{amsmath,amssymb,amsfonts}
\usepackage{verbatim}
\usepackage{algorithmic}
\usepackage{graphicx}
\graphicspath{{./figures/}}
\usepackage{textcomp}
\usepackage{xcolor}
\usepackage{bm}
\usepackage{subfigure}
\usepackage{makecell}
\usepackage{algorithm}
\def\BibTeX{{\rm B\kern-.05em{\sc i\kern-.025em b}\kern-.08em
    T\kern-.1667em\lower.7ex\hbox{E}\kern-.125emX}}

\begin{document}
\title{Few-Shot Specific Emitter Identification via Hybrid Data Augmentation and Deep Metric Learning}
\author{Cheng Wang, 
	Xue Fu,
	Yu Wang,
	Guan Gui, \emph{Senior Member}, \emph{IEEE},
	Haris Gacanin, \emph{Fellow}, \emph{IEEE},\\
	Hikmet Sari, \emph{Life Fellow}, \emph{IEEE},
	and Fumiyuki Adachi, \emph{Life Fellow}, \emph{IEEE}
\IEEEcompsocitemizethanks{
\IEEEcompsocthanksitem Cheng Wang, Xue Fu, Yu Wang, Guan Gui, and Himet Sari are with the College of Telecommunications and Information Engineering, Nanjing University of Posts and Telecommunications, Nanjing 210003, China (e-mail: 1021010517@njupt.edu.cn, 1020010415@njupt.edu.cn, 1018010407@njupt.edu.cn, guiguan@njupt.edu.cn, hsari@ieee.org).
\IEEEcompsocthanksitem Bamidele Adebisi is with the Department of Engineering, Faculty of Science and Engineering, Manchester Metropolitan University, Manchester M1 5GD, United Kingdom (e-mail: b.adebisi@mmu.ac.uk).
\IEEEcompsocthanksitem Haris Gacanin is with the Institute for Communication Technologies and Embedded Systems, RWTH Aachen University, Aachen 52062, Germany (e-mail: harisg@ice.rwth-aachen.de).
\IEEEcompsocthanksitem Fumiyuki Adachi is with the International Research Institute of Disaster Science (IRIDeS), Tohoku University, Sendai 980-8572 Japan (e-mail: adachi@ecei.tohoku.ac.jp).
}}

\markboth{IEEE Wireless Communications Letters,~Vol.~XX, No.~XX, XXX~2022}{}
\maketitle

\maketitle

\begin{abstract}
Specific emitter identification (SEI) is a potential physical layer authentication technology, which is one of the most critical complements of upper layer authentication. Radio frequency fingerprint (RFF)-based SEI is to distinguish one emitter from each other by immutable RF characteristics from electronic components. Due to the powerful ability of deep learning (DL) to extract hidden features and perform classification, it can extract highly separative features from massive signal samples, thus enabling SEI. Considering the condition of limited training samples, we propose a novel few-shot SEI (FS-SEI) method based on hybrid data augmentation and deep metric learning (HDA-DML) which gets rid of the dependence on auxiliary datasets. Specifically, HDA consisting rotation and CutMix is designed to increase data diversity, and DML is used to extract high discriminative semantic features. The proposed HDA-DML-based FS-SEI method is evaluated on an open source large-scale real-world automatic-dependent surveillance-broadcast (ADS-B) dataset and a real-world WiFi dataset. The simulation results of two datasets show that the proposed method achieves better identification performance and higher feature discriminability than five latest FS-SEI methods.
\end{abstract}

\begin{IEEEkeywords}
	Specific emitter identification (SEI), few-shot learning (FSL), data augmentation, deep metric learning.
\end{IEEEkeywords}

\section{Introduction}
In recent years, the rapid development of the Internet of Things (IoT) has accelerated the integration of many edge applications \cite{ref1,ref1a}. Thus, due to high density of IoT devices it is necessary to consider the secure IoT communications. Hence, it is important to conduct specific emitter identification (SEI) which can serve as the method of identification and certification \cite{ref2}. 
Due to the increased availability of big data and the increase in hardware computing power, deep learning (DL) and deep neural networks (DNN) for joint features extraction and classification have been successfully applied in many fields \cite{ref3,ref4,ref5,ref6,ref6a}. There are some DL-based SEI methods considered from the perspective of models, such as convolutional neural network (CNN)-based methods \cite{ref3} and recurrent neural network (RNN)-based methods \cite{ref7}, which achieve great performance. In addition, there are some methods using data in the transform domain as input of DNN to further improve identification performance, such as bispectrum \cite{ref8} and differential constellation trace figure (DCTF) \cite{ref9}.

DL-based SEI methods which are sufficient data-driven methods rely heavily on a large number of labeled samples to fully train DNN \cite{ref10}, while massive samples are not available in practical non-cooperative strong adversarial environments, making the identification performance of these methods decrease sharply due to insufficient training of the network. 
To overcome the limitations imposed by the dependence of DL-based SEI on massive samples, the study of SEI for few-shot (FS) scenarios has been considered. There are methods to study FS-SEI problem from the perspective of meta-learning \cite{ref11}, \cite{ref12}, deep metric learning (DML) \cite{ref13} and data augmentation \cite{ref14}, which achieve good identification performance. Although the above FS-SEI methods achieve better identification performance, these methods are still dependent on auxiliary datasets. These methods obtain a set of good initialization parameters from auxiliary dataset and then fine-tune the model parameters slightly on the target FS dataset. These auxiliary datasets are often extremely similar to the FS dataset, but it is often difficult to obtain such an auxiliary dataset in practical tasks. Liu $et \; al.$ \cite{ref20} and Cai $et \; al.$ \cite{ref21} considered the scene without auxiliary datasets, and they used adversarial training (AT) and virtual adversarial training (VAT) to achieve good identification performance, respectively, but the models of these two methods have lots of parameters which makes the models difficult to train.

In this paper, to overcome the dependence of DL-based SEI on plenty of training samples and get rid of reliance on auxiliary datasets, we propose a FS-SEI method based on Hybrid Data Augmentation and Deep Metric Learning (HDA-DML). Specifically, HDA is used to increase the quantity and diversity of training samples and improve the robustness of the model, while DML is used to extract high discriminative semantic features. The main contributions of this paper are summarized as follows:
\begin{itemize}
\item We propose a HDA-based FS-SEI method, in which rotation and CutMix augment the dataset in the data preprocessing and training process, respectively, and the DNN can learn more data distribution information from the augmented dataset.
\item We propose a DML-based FS-SEI method, in which triplet loss is used as the regularization term on the objective function to enable efficient learning in FS scenarios and improve the discriminability between inter-class semantic features.
\item We conduct the experiments to validate the proposed HDA-DML-based FS-SEI method on ADS-B and WiFi dataset. The experimental results show that the proposed method has the best identification performance and feature discriminability.
\end{itemize}

\section{Problem Formulation}
\begin{figure}[h]
	\centering
	\includegraphics[width=0.95\linewidth]{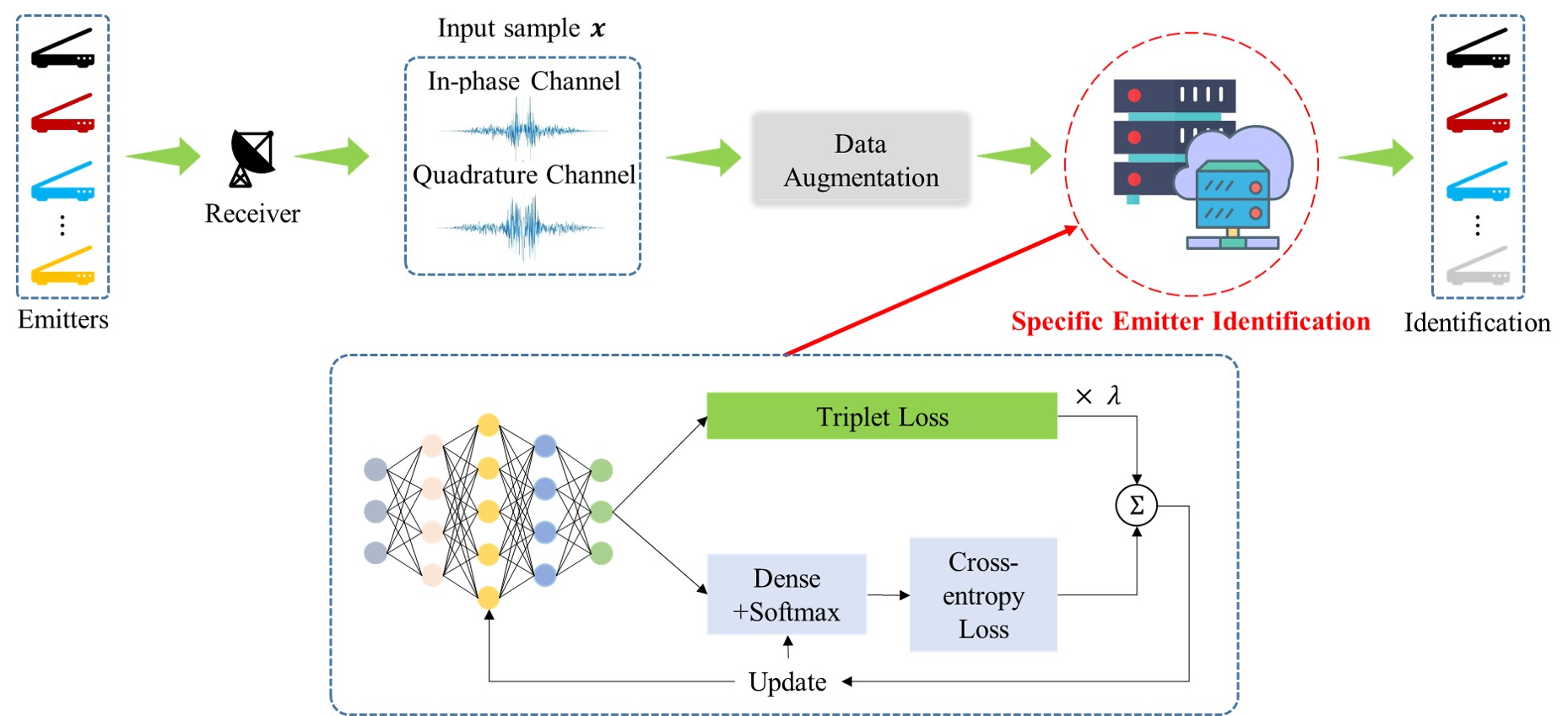}
	\caption{System model of HDA-DML-based FS-SEI.}
	\label{fig1}
\end{figure}

The system model of the proposed HDA-DML-based FS-SEI is shown in Fig. \ref{fig1}. The general three steps of system model can be described as: data collection, model training and identification.

Considering a general machine learning-based SEI problem, the goal of the problem is to find a maping function to minimize the expected error approximately, which can be written as:
\begin{equation}
	\underset{f \in \mathcal{F}}{\text{min}}\,{\varepsilon}_{em}=\underset{f \in \mathcal{F}}{\text{min}}\,\mathbb{E}_{(\bm{x},y) \in \bm{D}}\,\mathcal{L}(f(\bm{x}),y),
\end{equation}
where $\mathbb{E}$ denotes computing average, $f(\cdot)$ is the function that maps sample $\bm{x}$ to its predicted category, and $\mathcal{L}(\cdot)$ represents the object function that compares the predicted category $f(\bm{x})$ with the ground-truth category $y$. $\bm{x}$ represents the input sample with in-phase quadrature (IQ) format, and $y$ respresents the ground-truth category of the corresponding sample. We use $\bm{D}=\{(\bm{x}_{i},y_{i})\}_{i=1}^{N}$ to represent the dataset, in which $N$ is the number of samples, $\bm{x}_{i} \in \mathcal{X}$, $y_{i} \in \mathcal{Y}$, where $\mathcal{X}$ and $\mathcal{Y}$ represent sample space and category space, respectively. 

Different from the SEI problem, where the $\bm{D}$ consists of massive samples, the goal of FS-SEI problem without auxiliary dataset is to train an excellent mapping function $f(\cdot)$ using few samples. The problem can be described by few-shot dataset $\bm{D}_{fs}=\{\bm{D}_{tr},\bm{D}_{te}\}$, where $\bm{D}_{tr}=\{(\bm{x}_{i},y_{i})\}_{i=1}^{N_{tr}}$ is training dataset and $\bm{D}_{te}=\{(\bm{x}_{i},y_{i})\}_{i=1}^{N_{te}}$ is testing dataset. Specifically, there are $C$ categories with $K$ samples per category for $\bm{D}_{tr}$, and the total number of samples $N_{tr}$ in $\bm{D}_{tr}$ which is formulated as $N_{tr}=C \times K$ is usually small. Thus, it is denoted as a “C-ways, K-shots” problem, the goal of which is to use just $N_{tr}$ samples in $\bm{D}_{tr}$ to train an excellent mapping function $f(\cdot)$. Hence, it can be formulated so as to to perform the minimization below:
\begin{equation}
	\underset{f \in \mathcal{F}}{\text{min}}\,{\varepsilon}_{em}=\underset{f \in \mathcal{F}}{\text{min}}\,\mathbb{E}_{(\bm{x},y) \in \bm{D}_{tr}}\,\mathcal{L}(f(\bm{x}),y).
\end{equation}

\section{The Proposed HDA-DML-based FS-SEI Method}
\begin{figure}[h]
	\centering
	\includegraphics[width=0.95\linewidth]{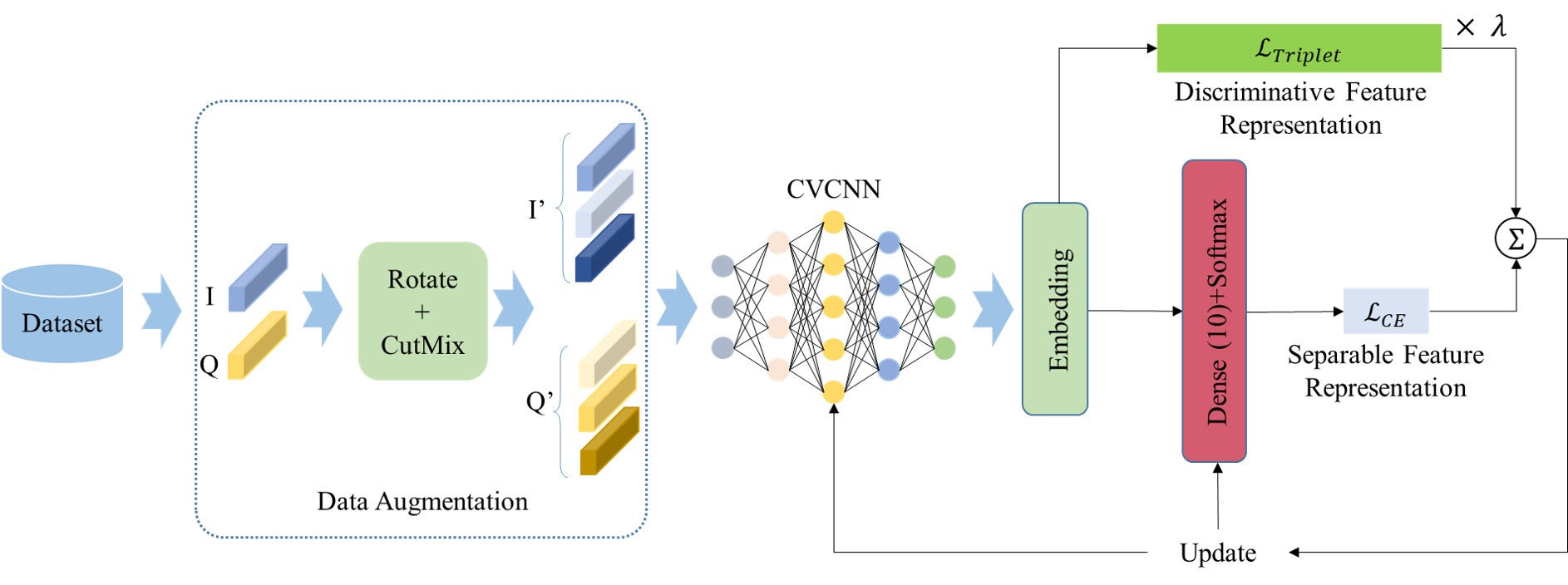}
	\caption{The framework of HDA-DML.}
	\label{fig2}
\end{figure}

The framework of HDA-DML is shown in Fig. \ref{fig2}. We use rotation and CutMix to extend training dataset, extract semantic features of the augmented samples via a complex-valued CNN (CVCNN) \cite{ref15} which can possess a more efficient and powerful feature extraction capability than CNN for complex signals containing coupling information, and optimize the CVCNN by using triplet loss as the regularization term on cross-entropy (CE) loss to extract separable and discriminative semantic features.

\subsection{The Data Augmentation of HDA-DML}
Data augmentation has wide application in DL-based SEI methods especially in FS-SEI methods, which can increase the diversity of training dataset, prevent model from over fitting and improve the robustness of the model. In this paper, two data augmentation methods, rotation and CutMix, are used to improve the diversity of training samples. These two methods are described as follows.
\subsubsection{Rotation \cite{ref14}}
For a sample point in an original signal sample $(I,Q)$, the augmented point $(I^{\prime},Q^{\prime})$ can be obtained by the following rotation transformation:
\begin{equation}
\left[
\begin{matrix}
	I^{\prime} \\
	Q^{\prime} \\
\end{matrix}
\right]
=\left[
\begin{matrix}
	\cos\alpha & -\sin\alpha \\
	\sin\alpha & \cos\alpha \\
\end{matrix}
\right]
\left[
\begin{matrix}
	I \\
	Q \\
\end{matrix}
\right],
\end{equation}
where $\alpha \in \{0, 0.5\pi, \pi, 1.5\pi\}$ is the angle of rotation. Rotation can extend one original signal sample into four signal samples.

\subsubsection{CutMix \cite{ref16}}
Assume that $\bm{x}$ and $y$ represent the input training samples and categories, respectively. The purpose of CutMix is to generate a series of new samples $(\tilde{\bm{x}},\tilde{y})$ by combining two original samples $(\bm{x}_A,y_A)$ and $(\bm{x}_B,y_B)$, both of which belong to the training dataset $\bm{D}_{tr}$, and then we feed the generated new samples into the CVCNN. Compared with other data augmentation, optimizing the CutMix-aided CVCNN not only makes full use of all information of the samples, but also uses mixed samples and mixed labels, which can fully consider the global distribution of data, so the optimized CVCNN will have better robustness towards unknown samples in testing process. Specifically, the new samples $(\tilde{\bm{x}},\tilde{y})$ can be generated through the following formulas:
\begin{equation}
	\tilde{\bm{x}}=\bm{M} \odot \bm{x}_A + (\bm{1}-\bm{M}) \odot \bm{x}_B,
\end{equation}
\begin{equation}
	\tilde{y}=\lambda y_A + (1-\lambda_{CM}) y_B,
\end{equation}
where $\bm{M}$ denotes a binary mask, the size of which is same as original sample, and it indicates where to delete and fill in two samples, and $\bm{1}$ represents a binary mask filled by ones, the size of which is also same as original sample. $\odot$ represents element-wise multiplication. $\lambda_{CM}$ is the combination ratio between two original samples, which is randomly selected from beta distribution $Beta(1,1)$. 

\subsection{The Objective Function Regularized with Metric}
In addition to increasing the diversity of FS samples, we also impose a regularization term on the objective function to enable efficient learning in FS scenarios. DML can learn a suitable distance to optimize the performance of the classifier and is an efficient approach for learning few samples \cite{ref24}. Compared with the network based only on CE loss, the network with metric loss can not only extract the separable semantic features, but also extract the discriminative semantic features \cite{ref25}. In this paper, to make the distance between similar semantic features (from the same category) closer and the distance between different semantic features (from the different categories) farther, the CVCNN in HDA-DML uses a triplet loss \cite{ref17} as regularization term of objective function to optimize feature extraction. The triplet loss function ${\mathcal{L}_{Triple}}$ can be formulated as:
\begin{equation}
	\mathcal{L}_{Triple }=\sum_{i=1}^{N}\left[d_{ap}-d_{an}+ \gamma\right]_{+},
\end{equation}
where $d_{ap}=\left\|g\left(\bm{x}_{i}^{a}\right)-g\left(\bm{x}_{i}^{p}\right)\right\|_{2}$ stands for the distance between $\bm{x}_{i}^{a}$ and $\bm{x}_{i}^{p}$, $d_{an}=\left\|g\left(\bm{x}_{i}^{a}\right)-g\left(\bm{x}_{i}^{n}\right)\right\|_{2}$ represents the distance between $\bm{x}_{i}^{a}$ and $\bm{x}_{i}^{n}$, and $g(\bm{x})$ represents the semantic feature of sample $\bm{x}$. $\bm{x}_{i}^{a}$ is a anchor sample which is randomly selected from any category, $\bm{x}_{i}^{p}$ is a positive sample which represents other samples of the same category as the anchor sample $\bm{x}_{i}^{a}$, and $\bm{x}_{i}^{n}$ is a negative sample which represents samples of different categories from the anchor sample $\bm{x}_{i}^{a}$. $||\cdot||_{2}$ indicates Euclidean distance, $\gamma$ is a margin which is an adjustable hyperparameter that can take on a positive value and $[\cdot]_+$ denotes positive part. 

In this paper, the triplet loss is used as the regularization term on the CE loss to optimize the CVCNN to extract the semantic features with both separability and high discriminability. The objective function can be written as follows:
\begin{equation}
	\mathcal{L}_{joint}=\mathcal{L}_{CE}+\lambda \mathcal{L}_{Triple},
\end{equation}
where $\mathcal{L}_{CE}$ represents the CE loss. The threshold $\lambda$ is used to balance the two loss functions. Specifically, the values of $\gamma$ and $\lambda$ are set with reference to \cite{ref13}.

\subsection{Training Prodedure}\label{AA} % \paragraph{Training Prodedure}
The total training procedure of HDA-DML is described in Algorithm \ref{alg1}. Adam optimizer is used for back propagation to update the parameters of CVCNN. In more detail, the rotation augmentation is operated in data preprocessing, while the CutMix augmentation is operated in the training process.
\begin{algorithm}[!ht]
	\renewcommand{\algorithmicrequire}{\textbf{Input:}}
	\renewcommand{\algorithmicensure}{\textbf{Output:}}
	\caption{Training procedure of the proposed HDA-DML-based FS-SEI method.}
	\textbf{Require:}
	\begin{itemize}
		\item $l_{r}$: learning rate
		\item $\theta$: the parameters of network
		\item $\bm{x}$, $\bm{x}^{\prime}$: raw samples and rotated samples
		\item $\bm{x}^{a}$, $\bm{x}^{p}$, $\bm{x}^{n}$: anchor, positive and negative samples
		\item $\tilde{\bm{x}}$, $\tilde{y}$: sample and label after CutMix
		\item $\bm{M}$: binary mask of CutMix
		\item $\lambda_{CM}$: combination ratio of CutMix
		\item $\mathcal{L}_{CE}$, $\mathcal{L}_{Triple}$, $\mathcal{L}_{joint}$: CE loss, triplet loss and joint loss
		\item $\lambda$: threshold between CE loss and triplet loss
		\item $T$: the number of training iterations
		\item $B$: the number of batches in a training iteration
	\end{itemize}
	\begin{algorithmic}[1]
		\REQUIRE Few-shot training dataset $\bm{D}_{tr}=\{(\bm{x}_i,y_i)\}_{i=1}^{N_{tr}}$
		\ENSURE Predicted categories $\hat{y}$
    	\STATE Data preprocessing:
    	\begin{itemize}
    		\item Rotation: $\bm{x}^{\prime} \leftarrow \bm{x}$
    		\item Power normalization: $\bm{x}^{\prime}\leftarrow\frac{\bm{x}^{\prime}-\bm{x}^{\prime}_{\text{min}}}{\bm{x}^{\prime}_{\text{max}}-\bm{x}^{\prime}_{\text{min}}}$
    	\end{itemize}
    	\STATE Building network and randomly initializing $\theta$
    	\FOR{$t = 1$ to $T$}
    	\FOR{$b = 1$ to $B$}
    	\STATE $\tilde{\bm{x}}^{\prime}=\bm{M} \odot \bm{x}^{\prime}_{i} + \bm{(1-M)} \odot \bm{x}^{\prime}_{j}$
    	\STATE $\tilde{y}=\lambda_{CM} y_{i} + (1-\lambda_{CM}) y_{j}$
    	\STATE Feeding the augmented signals $\{\tilde{\bm{x}}^{\prime},\tilde{y}\}$ into CVCNN
    	\STATE Extracting semantic features $g(\tilde{\bm{x}}^{\prime})$
    	\STATE Calculating the predicted categories $f(g(\tilde{\bm{x}}^{\prime}))$
    	\STATE Computing objective loss function: $\mathcal{L}_{joint}(\tilde{\bm{x}}^{\prime},\tilde{y})=\mathcal{L}_{CE}(f(g(\tilde{\bm{x}}^{\prime})),\tilde{y})+\lambda\mathcal{L}_{Triple}(g(\tilde{\bm{x}}^{a\prime}),g(\tilde{\bm{x}}^{p\prime}),g(\tilde{\bm{x}}^{n\prime}))$
    	\STATE Updating parameters of CVCNN with backward propagation: $\theta \leftarrow \text{Adam}(\nabla_{\theta},\mathcal{L}_{joint},l_{r},\theta)$
    	\ENDFOR
    	\ENDFOR
    \end{algorithmic}
\label{alg1}
\end{algorithm}

\section{Experimental Results}
\subsection{Simulation Parameters}
In this paper, we use two different datasets to evaluate HDA-DML-based FS-SEI method. The first dataset that was presented in \cite{ref18} contains ADS-B signals collected in a real-world large-scale airspace and is suitable for SEI researches. Specifically, we randomly select 10 categories of long IQ signals, the number of sampling points is 6000 and the signal-to-noise ratio (SNR) is 30 dB. The second dataset presented in \cite{ref19} contains WiFi signals collected from 16 X310 USRP devices, the SNR of which is also 30 dB and we cut them into signal samples with 6000 sampling points.

We build five few-shot experimental scenarios to evaluate the identification performance of HDA-DML-based FS-SEI method. For both datasets, each scenario is \{1, 5, 10, 15, 20\} shots. TABLE \ref{tab3} shows the details of other experimental parameters.
\begin{table}[htbp]
	\caption{Detailed simulation parameters.}
	\begin{center}
		\resizebox{0.45\textwidth}{!}{
		\begin{tabular}{|c|c|}
			\hline
			Items & Parameters \\
			\hline
			Margin $\gamma$ of Triplet Loss & 5 \\
			\hline
			Threshold $\lambda$ & 0.01 \\
			\hline
			Optimizer & $\text{Adam}$ \\
			\hline
			Batch Size & 16 \\
			\hline
			Learning Rate $l_{r}$ & 0.01 \\
			\hline
			Simulation Platforms & Pytorch NVIDIA GeForce GTX 1080Ti \\
			\hline
		\end{tabular}}
		\label{tab3}
	\end{center}
\end{table}

\subsection{Evaluation Criteria and Benchmarks}
In this paper, we use identification accuracy and silhouette coefficient to evaluate identification performance and semantic feature discriminability, respectively. 
The identification accuracy can be expressed as the ratio between the number of correctly identified test samples and the total number of test samples.
The silhouette coefficient can measure the cohesion and separation degree of the extracted semantic features, and we use it as an indicator of discriminability between features.

In this paper, we compare the HDA-DML method with three latest FS-SEI methods, namely, Triplet-CVCNN \cite{ref13}, CRCN-AT \cite{ref20} and VAT \cite{ref21}. In addition, HDA-DML method is also compared with Softmax-CVCNN \cite{ref22} and DA-CVCNN \cite{ref14}. Taking fairness into consideration, without changing the core idea of these methods, we use the same training samples with IQ format, the same data preprocessing method, optimizer, learning rate and network structure.

\subsection{Identification Performance Comparison}
\subsubsection{HDA-DML $vs.$ Benchmarks}
\begin{figure}[h]
	\centering
	\includegraphics[width=1\linewidth]{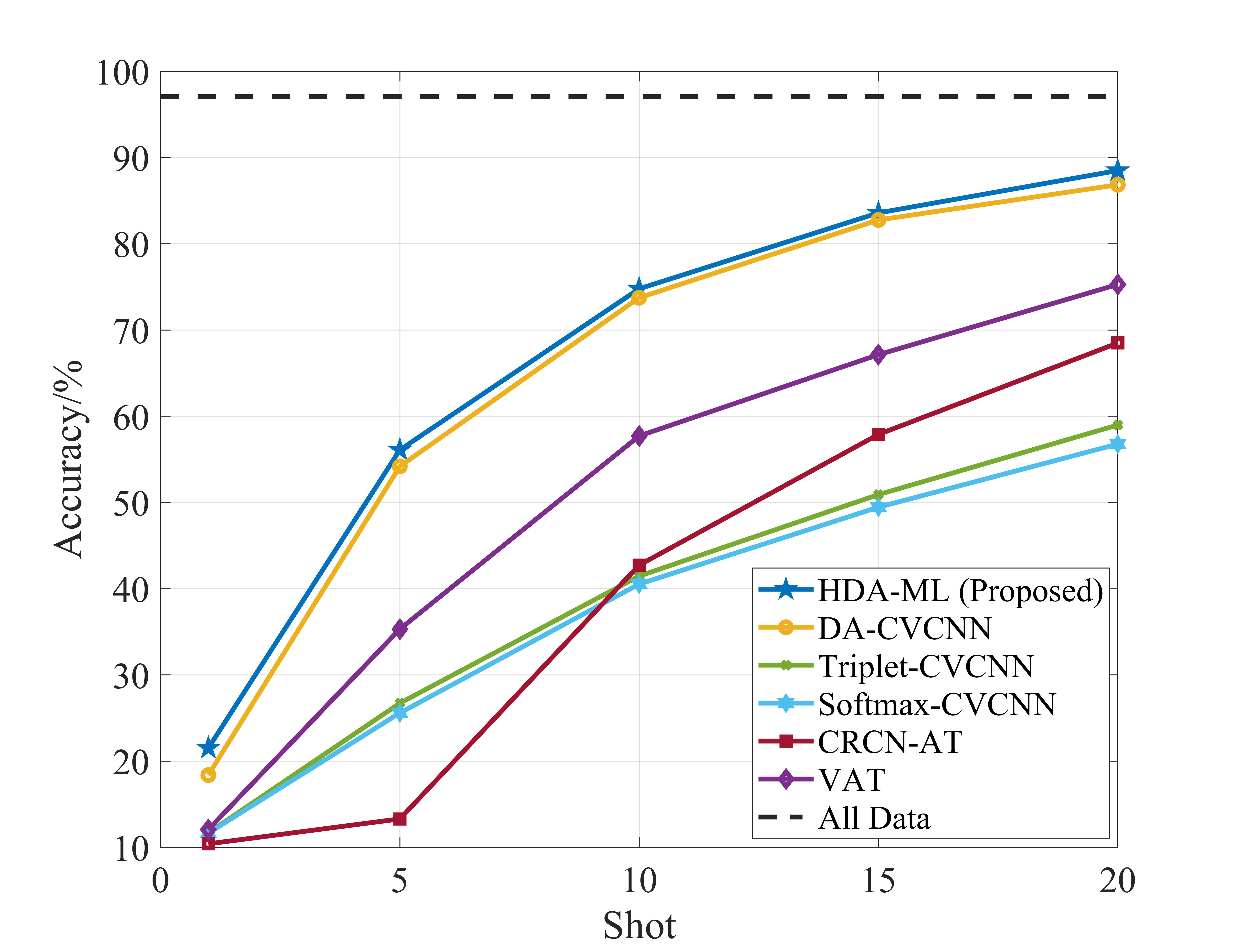}
	\caption{The identification performance of HDA-DML-based FS-SEI method and benchmarks in ADS-B dataset, where ``All Data'' represents the identification accuracy on the test dataset after training with 2611 samples.}
	\label{fig4}
\end{figure}

\begin{figure}[h]
	\centering
	\includegraphics[width=1\linewidth]{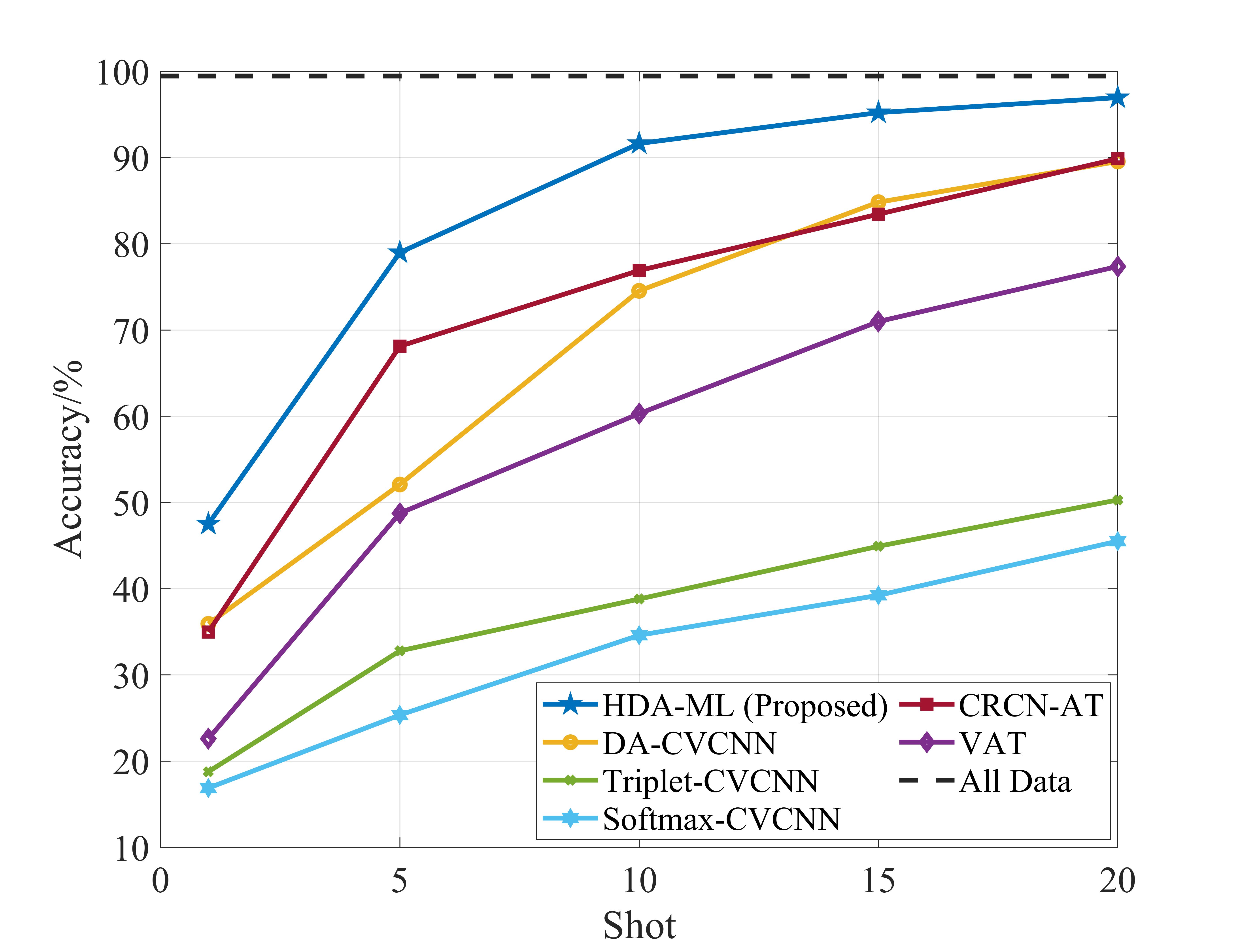}
	\caption{The identification performance of HDA-DML-based FS-SEI method and benchmarks in WiFi dataset, where ``All Data'' represents the identification accuracy on the test dataset after training with 34140 samples.}
	\label{fig5}
\end{figure}

Due to the instability of the sample quality, the experimental results shown are the average of the results of 100 Monte Carlo simulations. The identification performance of HDA-DML-based FS-SEI method and comparative methods in ADS-B dataset and WiFi dataset are shown in Fig. \ref{fig4} and Fig. \ref{fig5}, respectively. It can be seen that our proposed method has a clear improvement in identification performance compared to the comparison methods for all few-shot scenarios in both datasets. This indicates that the robustness of our proposed method is better. Specifically, compared with the comparison methods, in ADS-B dataset, the identification accuracy of HDA-DML can be improved by at least 3$\%$, and in WiFi dataset, the identification accuracy can be improved by at least 10$\%$.

\subsubsection{Visualization of Semantic Features}
\begin{figure}[h]
	\centering
	\includegraphics[width=1\linewidth]{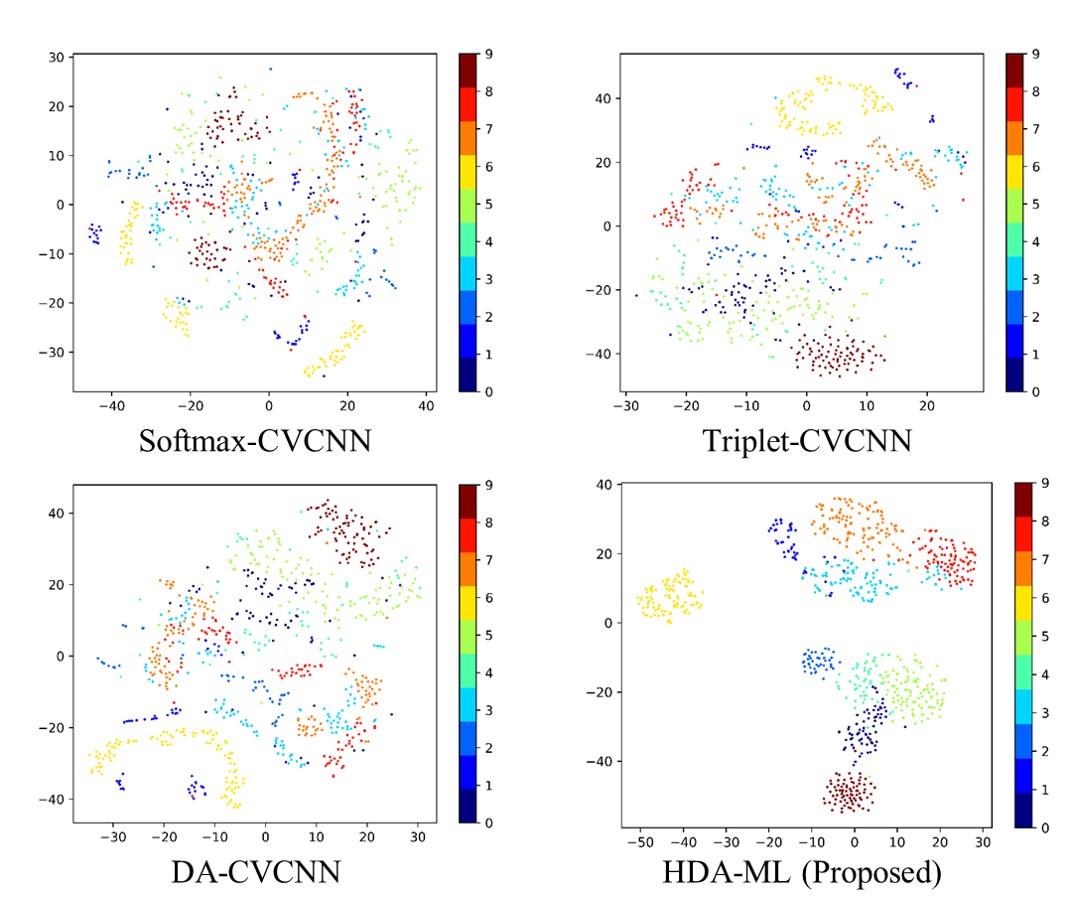}
	\caption{Visualization of semantic features in ADS-B dataset.}
	\label{fig6}
\end{figure}

The silhouette coefficients of HDA-DML and the comparison methods on the two datasets in ``20 shots'' scenario are shown in TABLE \ref{tab4}. As can be seen from the table, on both datasets, HDA-DML is able to achieve the best silhouette coefficient and the discriminability of semantic features is the best.
\begin{table*}[htbp]
	\caption{Silhouette coefficients of HDA-DML and comparison methods.}
	\begin{center}
		\resizebox{0.9\textwidth}{!}{
			\begin{tabular}{|c|c|c|c|c|c|c|}
				\hline
				Datasets & {\bf HDA-DML (our proposed)}  & CRCN-AT &   VAT    & Triplet-CVCNN & DA-CVCNN &  Softmax-CVCNN  \\
				\hline
				ADS-B & $\bm{0.46245}$ & 0.00047 & $-0.01991$ & 0.10381 & 0.044798 & $-0.02991$ \\
				\hline
				WiFi  & $\bm{0.69685}$ & 0.69077 &  0.61927 & 0.14791 & 0.601797 & 0.16032 \\
				\hline
		\end{tabular}}
		\label{tab4}
	\end{center}
\end{table*}
In this paper, we use t-distributed stochastic neighbor embedding (t-SNE) \cite{ref23} for feature compression, and the visualization figures of the compressed semantic features are shown in Fig. \ref{fig6}. Due to the limitation of space, we only give the figures of ADS-B dataset.

It can be clearly seen that Softmax-CVCNN only uses the CE loss function, and the separability between different categories is weak; Triplet-CVCNN has smaller intra-class distance and larger inter-class distance compared with Softmax-CVCNN because it uses the triplet loss, but the extracted semantic features of Softmax-CVCNN and Triplet-CVCNN are haphazardly distributed throughout the feature space and it is difficult to find clear boundaries between different categories of samples because these methods can not learn the data information efficiently from few samples. Since data augmentation can extend the diversity of samples, the features extracted by DA-CVCNN have better separability but still weak discriminability, the semantic features extracted by our proposed HDA-DML method have the best separability and discriminability, which are significantly better than the other compared methods.

\section{Conclusion}
We proposed an effective FS-SEI method based on HDA-DML. The proposed method considers FS-SEI without auxiliary datasets and innovatively combines two types of data augmentation, not only to achieve data expansion but also to consider the global features of the samples, while using the triplet loss as the regularization of objective function to optimize the network and extract discriminative semantic features. We validated the performance of this method on the ADS-B and WiFi dataset. Without the use of auxiliary datasets, the proposed method achieves a great improvement in identification performance and feature discriminability compared to the comparison methods.

\end{document}